\documentclass[journal=jpccck,manuscript=letter,layout=twocolumn]{achemso}

\usepackage[version=3]{mhchem} 
\usepackage{graphicx}
\usepackage{dcolumn}
\usepackage{bm}
\usepackage[mathlines]{lineno}
\usepackage{amssymb}
\usepackage{amsmath}
\usepackage{soul,color}
\usepackage[colorlinks,linkcolor=blue,anchorcolor=blue,citecolor=blue,urlcolor=blue]{hyperref}
\usepackage{dcolumn}
\usepackage{threeparttable}  
\usepackage{cmll}
\usepackage{multirow}
\usepackage{booktabs}
\usepackage{color}
\usepackage{datetime}

\title{Highly Anisotropic Electronic and Mechanical Properties of 
       Monolayer and Bilayer As$_2$S$_3$}

\author{Xuefei Liu}
\affiliation{Key Laboratory of Low Dimensional Condensed Matter Physics
             of Higher Educational Institution of Guizhou Province, %
             Guizhou Normal University, Guiyang 550025, China}
\alsoaffiliation{Low-Dimensional Semiconductor Structure Laboratory, %
                 College of  Big data and Information Engineering, %
                 Guizhou University, Guiyang 550025, China}

\author{Zhaofu Zhang}
\affiliation{Department of Engineering, University of Cambridge, Cambridge, CB2 1PZ, United Kingdom}

\author{Zhao Ding}
\affiliation{Low-Dimensional Semiconductor Structure Laboratory, %
                 College of  Big data and Information Engineering, %
                 Guizhou University, Guiyang 550025, China}

\author{Bing Lv}
\affiliation{Key Laboratory of Low Dimensional Condensed Matter Physics
             of Higher Educational Institution of Guizhou Province, %
             Guizhou Normal University, Guiyang 550025, China}

\author{Zijiang Luo}
\affiliation{College of Information, Guizhou Finance and Economics University, Guiyang 550025, China}

\author{Jian-Sheng Wang}
\affiliation{Department of Physics, National University of Singapore,
             Singapore 117551, Republic of Singapore}

\author{Zhibin Gao}
\email{zhibin.gao@nus.edu.sg}
\affiliation{Department of Physics, National University of Singapore,
             Singapore 117551, Republic of Singapore}

\date{\today}


\begin{document}


\begin{abstract}
Anisotropic materials, with orientation-dependent properties, %
have attracted more and more attention due to their compelling
tunable and flexible performance in electronic and optomechanical
devices. So far, two-dimensional (2D) black phosphorus shows the
largest known anisotropic behavior, which is highly desired for
synaptic and neuromorphic devices, multifunctional directional
memories, and even polarization-sensitive photodetector, whereas
it is unstable at ambient conditions. Recently, 2D few-layered
As$_2$S$_3$ with superior chemical stability was successfully
exfoliated in experiments~\cite{kins2019highly}. However, %
the electronic and mechanical properties of monolayer and
bilayer As$_2$S$_3$ is still lacking. Here, we report the large
anisotropic electronic and mechanical properties of As$_2$S$_3$
systems through first-principles calculations and general
angle-dependent Hooke's law. Monolayer and bilayer As$_2$S$_3$
exhibit anisotropic factors of Young's modulus of 3.15 and
3.32, respectively, which are larger than the black
phosphorous with experimentally confirmed and an anisotropic
factor of 2~\cite{tao2015mechanical}. This study provides
an effective route to flexible orientation-dependent
nanoelectronics, nanomechanics, and offers implications in
promoting related experimental investigations. %
\end{abstract}

\section*{Introduction}
A material is isotropic if its mechanical and elastic properties
are the same in all directions. When this is not correct, the
material is anisotropic. Many materials are anisotropic and even
inhomogeneous owning to the tunable formation and composition of
their constituents and elements. So far, most of
two-dimensional (2D) materials are isotropic, such as
distinguished graphene~\cite{novoselov2004electric,novoselov2005two}, %
h-BN~\cite{lin2012advances}, transition metal
dichalcogenides~\cite{wang2012electronics,gao2019degenerately}. %
There exist only a few anisotropic 2D crystals at present, such
as renowned black phosphorus~\cite{liu2014phosphorene}, %
SnSe~\cite{chang20183d}, and atomically thin
tellurium~\cite{zhu2017multivalency,gao2018unusually}. Since the
properties of the isotropic materials are the same in any
orientation, their behavior is therefore highly predictable. %
Most glasses and polymers are examples of isotropic materials, %
which have been widely used in the packaging industry, medical
equipment, and even home tableware. On the contrary, the properties
of anisotropic materials are direction-dependent, which usually
consists of asymmetric crystalline structures. Furthermore, %
artificial anisotropic single crystals (metamaterials) are also
highly desired as developing technology, such as selective
fluorescence DNA sensors~\cite{tan2015high}, anisotropic synaptic
devices for neuromorphic applications~\cite{tian2016anisotropic}, %
anisotropic nanoelectronics with multifunctional directional
memories in the 2D limit~\cite{wang2019gate}, digital
inverters~\cite{liu2015integrated}, and even polarization-sensitive
broadband photodetectors~\cite{yuan2015polarization}. %

Recently, low-symmetry 2D materials have attracted more and more
attention owing to the unique orientation-dependent properties
that are not easily obtained in the usual isotropic and symmetric
2D materials~\cite{sa2019elastic,zhao2020plane,shen2018resolving,%
yang2017optical,zhou2018highly,tao2015mechanical}. In these
anisotropic 2D materials, the electronic, optical, thermal, %
piezoelectric, and even ferroelectric properties are
direction-dependent, which would open up a new degree of freedom
to selectively tune the physical properties of 2D materials-based %
nano-devices~\cite{ali2014large,liu2015integrated,%
qiu2019thermoelectric,wang2019gate,yin2020anisotropic}. At present, %
2D black phosphorus (BP) shows the largest known anisotropy of Young's
modulus with a ratio of $\frac{E_b}{E_a} = 2$ along the in-plane
axes (\textit{b} and \textit{a})~\cite{tao2015mechanical,%
wei2014superior}. However, BP has a fatal disadvantage that is
unstable in the ambient conditions, which severely constrains
its potential applications~\cite{castellanos2014isolation}.

Recently, 2D few-layered As$_2$S$_3$ with superior chemical
stability was successfully exfoliated in the experiment by
$\check{S}$i$\check{s}$kins \textit{et al.}~\cite{kins2019highly}
and they have also systematically studied the anisotropic optical
properties such as Raman spectroscopy, resonance frequency
analysis using laser interferometry~\cite{kins2019highly}. %
However, the electronic and mechanical properties of monolayer
and bilayer As$_2$S$_3$ is still lacking. Is the anisotropy
of a single (bilayer) layer greater than that of a multi-layer
in As$_2$S$_3$? How does the anisotropy of monolayer and
bilayer As$_2$S$_3$ compare to the well-known BP? %

Here, the anisotropic electronic and mechanical properties of
monolayer and bilayer As$_2$S$_3$ are systematically studied
using first-principles methods combined with the
orientation-dependent 2D-plane Hooke's law. Furthermore, we
have elaborately analyzed the angle-resolved effective mass
of holes and electrons, angle-resolved Young's modulus,
Poisson's ratio and Shear modulus of monolayer and bilayer
As$_2$S$_3$. The calculated anisotropic factor of monolayer
and bilayer As$_2$S$_3$ are 3.15 and 3.32 respectively, %
which are quite larger than the renowned BP with an
anisotropic ratio of 2. Our studies will provide a more
comprehensive understanding and insights into the
potential applications of 2D As$_2$S$_3$ in
orientation-dependent nanoscience and nanotechnology. %

\section*{Computational methods}
%
\begin{figure*}[t!]
\includegraphics[width=2.0\columnwidth]{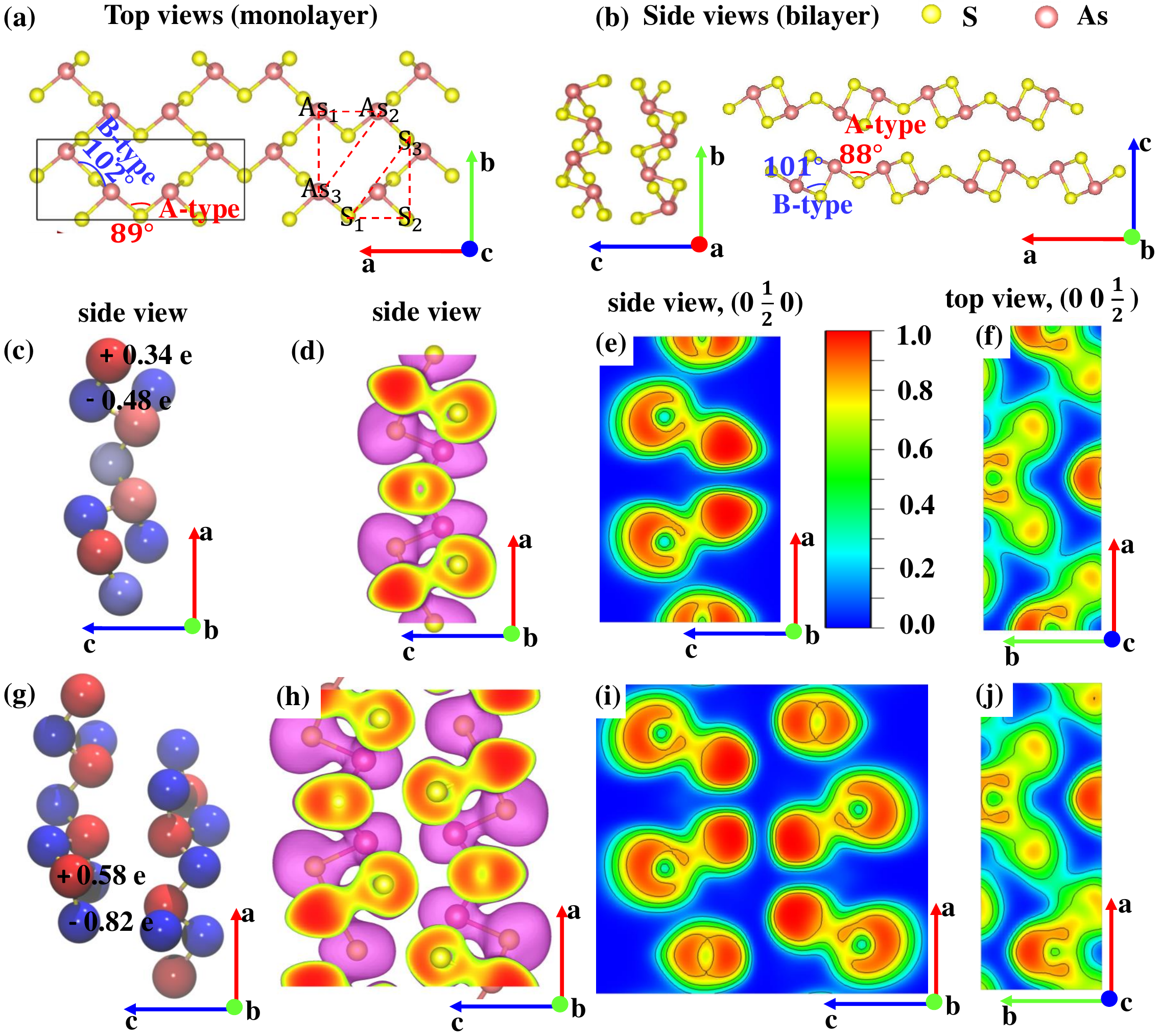}
\caption{(a)(b) Top and side views of the monolayer and bilayer
2D As$_2$S$_3$ in a 2 $\times$ 2 supercell. The primitive
cell of monolayer As$_2$S$_3$ is indicated by a solid black
rectangle. According to the different hinged deformation, the
As-S-As bond angles can be classified into two types (A and
B), shown in (a) and (b). (c) and (g) depict the Bader
charge distribution for monolayer and bilayer As$_2$S$_3$, %
respectively. (d)-(j) describe the electron localization
function (ELF) for monolayer (above) and bilayer (below) %
As$_2$S$_3$, separately. (d) and (h) are the 3D ELF and
the iso-surface value in the side views of ELF is 0.59. %
The Miller indexes for top views and side views are
(0 0 1/2) and (0 1/2 0). The blue and red colors in (c)
and (g) represent gaining and losing electrons, and
the number of transferred electrons are characterised
by the depth of colors. The maximum value of obtaining
and losing electrons are shown in (c) and (g). %
}\label{fig1}
\end{figure*}

The optimization of structures and static self-consistent energy calculations
were conducted using the Perdew-Burke-Ernzerhof (PBE) exchange-correlation
functional~\cite{perdew1996generalized} along with the projector-augmented
wave(PAW) potentials~\cite{blochl1994projector,kresse1999ultrasoft} as
implemented in the Vienna Ab-initio Simulation
Package (VASP)~\cite{kresse1996efficient,kresse1996efficiency}. The kinetic
energy cutoff was 400~eV and the linear tetrahedron method with Bl{\"o}chl
corrections~\cite{blochl1994improved} was used to integrate the Brillouin
zone. The reciprocal space was sampled with a $4 \times 12 \times 1$\ %
Monkhorst-Pack \textit{k}-point. All atoms were relaxed using the conjugate
gradient method until the Hellman-Feynman forces on individual atoms were
less than 0.02~eV/\AA ~and the total energy difference between two successive
steps were lower than $10^{-5}$~eV. To describe correctly the Van der Waals
iteration resulting from dynamical correlations between fluctuating charge
distributions, we adopted the DFT-D2 method of
Grimme~\cite{grimme2006semiempirical}, a correction to the conventional
Kohn-Sham DFT energy. A vacuum thickness with 20~\AA~was used to avoid the
fictitious interaction between adjacent images normal to the in-plane
direction. Since the PBE approach usually underestimates the band gap of
materials, we also adopted the screened hybrid functional of Heyd, Scuseria, %
and Ernzerhof (HSE06)~\cite{heyd2003hybrid} for a more accurate calculation
on the electronic band structures of As$_2$S$_3$ systems. The density functional
perturbation theory (DFPT) was used in a $3\times 5\times 1$ supercell. The
phonon dispersion was obtained using the Phonopy~\cite{togo2008first}. The
angle-resolved effective mass of holes and electrons, angle-resolved mechanical
properties were performed using VASPKIT code~\cite{wang2019vaspkit}. %

\section*{Results and discussions}
\subsection{Crystal structure and anisotropic charge distribution}
The optimized crystal structures of 2D monolayer and bilayer
As$_2$S$_3$ in a 2 $\times$ 2 supercell are shown in
Figure~\ref{fig1}. Monolayer As$_2$S$_3$ belongs to the
Orthorhombic crystal system with \textit{Pmn}2$_1$ symmetry
group (space group No.~31). The primitive cell of monolayer
As$_2$S$_3$ contains 10 atoms, highlighted by a black solid
rectangle in Figure~\ref{fig1}(a). There are two distinct
types of elements in monolayer and bilayer As$_2$S$_3$. The
hinged S atoms have a coordination number of two, while
the rigid As atoms have a coordination number of three. Due
to the weak Van der Waals interaction between layers, As$_2$S$_3$
can be exfoliated from the bulk counterpart, resulting in a
monolayer or layered As$_2$S$_3$ Membranes~\cite{kins2019highly}. %
Therefore, we would like to explore the bilayer As$_2$S$_3$ which
consists of 20 atoms. Note that the bilayer As$_2$S$_3$ remains
the same stacking formation with the 3D As$_2$S$_3$ phase
belonging to the \textit{P}$2_1/c$~\cite{mortazavi20202}, %
indicating the bilayer can be easily obtained from the
3D As$_2$S$_3$. The optimal interlayer intrinsic distance is
2.04 {\AA} based on the DFT-D2 functional, which is a
little smaller than the 3D As$_2$S$_3$ of 2.79 {\AA}. The
optimized lattice constants of the monolayer (bilayer) %
As$_2$S$_3$ are $\left| \vec{a} \right| =$ 11.42 (11.49) %
{\AA} and $\left| \vec{b} \right| =$ 4.41 (4.33) {\AA}, %
respectively, which are in a good agreement with the
previous work~\cite{mortazavi20202}. %

Bader charge is an effective analysis of assigning electron
density of molecules and solids to individual
atoms~\cite{henkelman2006fast}. The calculated Bader charge
of monolayer and bilayer As$_2$S$_3$ are shown in
Figs.~\ref{fig1}(c) and Figs.~\ref{fig1}(g), respectively. %
For comparison, red (blue) color represent losing (gaining) %
electrons, and the concentration of each color is used to
describe the amount of electrons charge transfer. Based on
our calculation, each arsenic atom loses an average of
0.34 (0.58) electrons for monolayer and bilayer As$_2$S$_3$, %
respectively. The different amount of electrons transfer
originates from the Van der Waals interaction between the
two sub-layers in the bilayer structure.

Electron localization function (ELF) is a measure of the
possibility of finding an electron in the neighborhood
space of a reference electron~\cite{silvi1994classification}. %
It is a three-dimensional function with a value ranging from
0 that indicates a low electron density localization and
metallic ionic bonds to 1 that implies strong covalent
bonding or lone pair electrons. The calculated ELF for
monolayer As$_2$S$_3$ are shown in Figs.~\ref{fig1}(e) and
\ref{fig1}(f). The result shows that ELF of sulfur atoms
is larger than that of arsenic atoms, indicating a large
anisotropic charge distribution of monolayer As$_2$S$_3$. %
Figs.~\ref{fig1}(i) and \ref{fig1}(j) also depict a similar
charge distribution of bilayer As$_2$S$_3$ with the monolayer
As$_2$S$_3$. Furthermore, the strong electron localization
locates between sulfur and arsenic atoms, indicating the
dominant role of covalent bonding in both monolayer and
bilayer As$_2$S$_3$.

\subsection{Anisotropic electronic transport properties}
%
\begin{figure*}[t!]
\includegraphics[width=2.0\columnwidth]{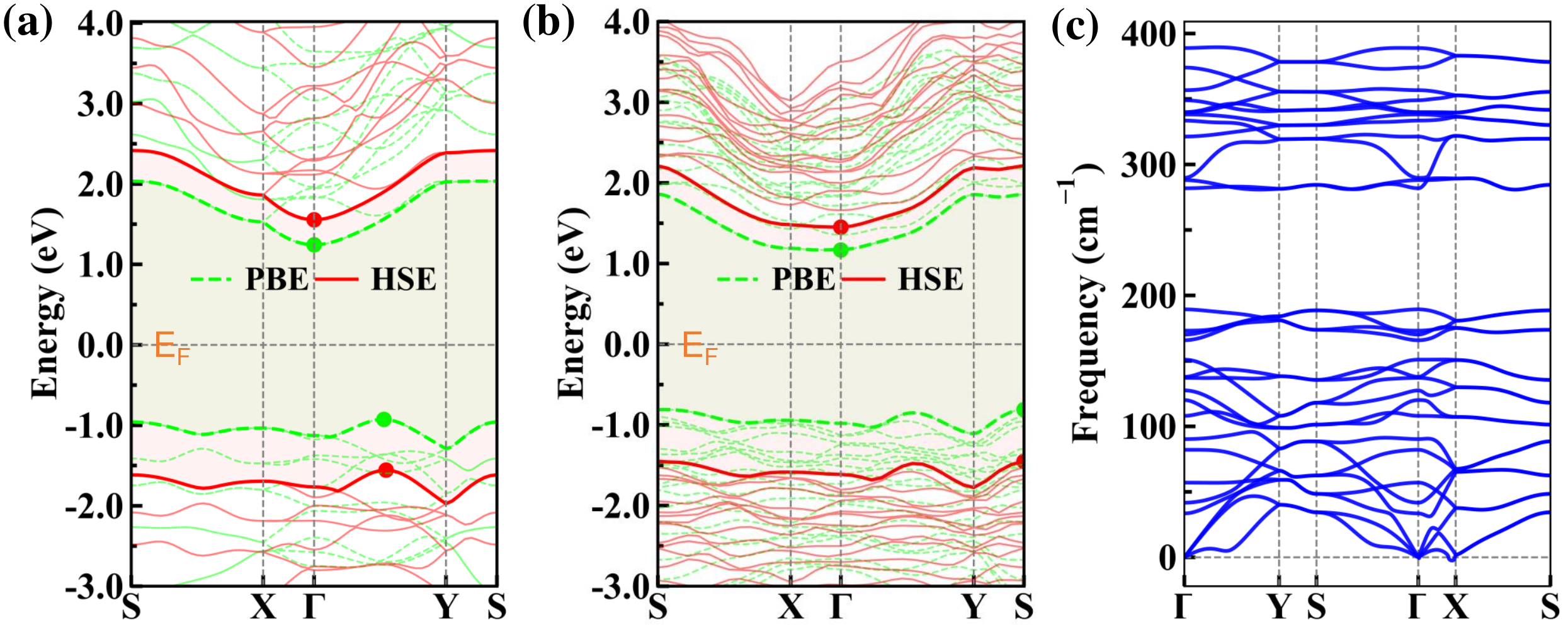}
\caption{Electronic band structures of (a) monolayer and (b)
bilayer 2D As$_2$S$_3$ using the DFT--PBE (dashed green) and
DFT--HSE06 (solid red) functionals.  In the first Brillouin
zone, the high symmetry \textit{k} points are: S(1/2 1/2 0), %
X(1/2 0 0), $\Gamma$(0 0 0), and Y(0 1/2 0), respectively. %
The Fermi levels are indicated by black dashed lines. The
valence band maximum (VBM) and conduction band minimum (CBM)
are also marked by colored balls. (c) Phonon dispersion of
monolayer As$_2$S$_3$. %
}\label{fig2}
\end{figure*}

We find that Bader charge and ELF can be further explained
by the different electronegativity of sulfur (2.58) and
arsenic (2.18) atoms. Therefore, sulfur atoms tend to gain
more electrons than arsenic atoms due to a larger
electronegativity. This covalent bond mechanism in bilayer
As$_2$S$_3$ is stronger than the monolayer As$_2$S$_3$, which
can be verified in Figs.~\ref{fig1}(e) and \ref{fig1}(i). %
What is more, Figs.~\ref{fig1}(f) and \ref{fig1}(j) display
the different electron densities in the 2D plane, especially
along $\vec{a}$ direction, no matter in monolayer and bilayer
As$_2$S$_3$. This outcome demonstrates that electrons are more
continuous and denser in the $\vec{a}$ axis compared with
the $\vec{b}$ axis, suggesting a large anisotropic charge
distribution for monolayer and bilayer As$_2$S$_3$. This
special electron behavior will lead to the anisotropic
properties of As$_2$S$_3$. We will discuss it in the
following sections. %

The calculated phonon dispersion of monolayer As$_2$S$_3$ is
shown in Figure~\ref{fig2}c which is free from the imaginary
frequencies, indicating the dynamical stability of As$_2$S$_3$. %
As expected, the phonon dispersion of As$_2$S$_3$ shows a typical
characteristic of 2D materials. It has three acoustic phonon
modes, in which two of three (TA and LA) show linear functions
around the $\Gamma$ point and the ZA mode is a quadratic
relation. Our phonon dispersion is consistent with the previous
result~\cite{mortazavi20202}. %

Next, we study the electronic properties of both monolayer and
bilayer As$_2$S$_3$. The calculated band structures are shown
in Figs.~\ref{fig2}(a) and \ref{fig1}(b), indicating indirect
semiconductors for monolayer and bilayer As$_2$S$_3$. For
monolayer, the conduction band minimum is located at the $\Gamma$
point while the valence band maximum (VBM) lies between the
$\Gamma$ and Y point (0 1/2 0). In the case of the bilayer, the
CBM changes a little but the CBM is transformed to the S
point (1/2 1/2 0). Based on the DFT-PBE calculations, the
monolayer As$_2$S$_3$ has an indirect band gap of 2.17~eV, and
the bilayer is 1.97~eV. This is consistent with the physical
picture that usually monolayer material has a larger band gap
than that of the few-layer material~\cite{gao2018high}. %

Since the fundamental band gap is usually underestimated in
DFT-PBE calculations, we have resorted to the HSE06. The
calculated HSE06 band gaps for monolayer and bilayer As$_2$S$_3$
are 3.11 and 2.91 eV, respectively. Our monolayer HSE06
band gap is 0.16 eV smaller than the previous
work~\cite{mortazavi20202} since we used a much dense \textit{k}
point to do the calculation. Note that the locations of CBM
and VBM are the same as their results~\cite{mortazavi20202}. It
is also found that the HSE06 method does not change not only the
shape of the band structures of monolayer and bilayer As$_2$S$_3$, %
but also for the positions of VBM and CBM (band edges). Besides, %
the bands shown in Figs.~\ref{fig2}(a) and \ref{fig1}(b) along
$\Gamma$--X direction are more non-dispersive than that of
$\Gamma$--Y direction, indicating strong anisotropy for monolayer
and bilayer As$_2$S$_3$. Therefore, we need more quantitative study
of electronic transport properties for As$_2$S$_3$ systems in the
following. %

\begin{figure*}[t!]
\includegraphics[width=2.0\columnwidth]{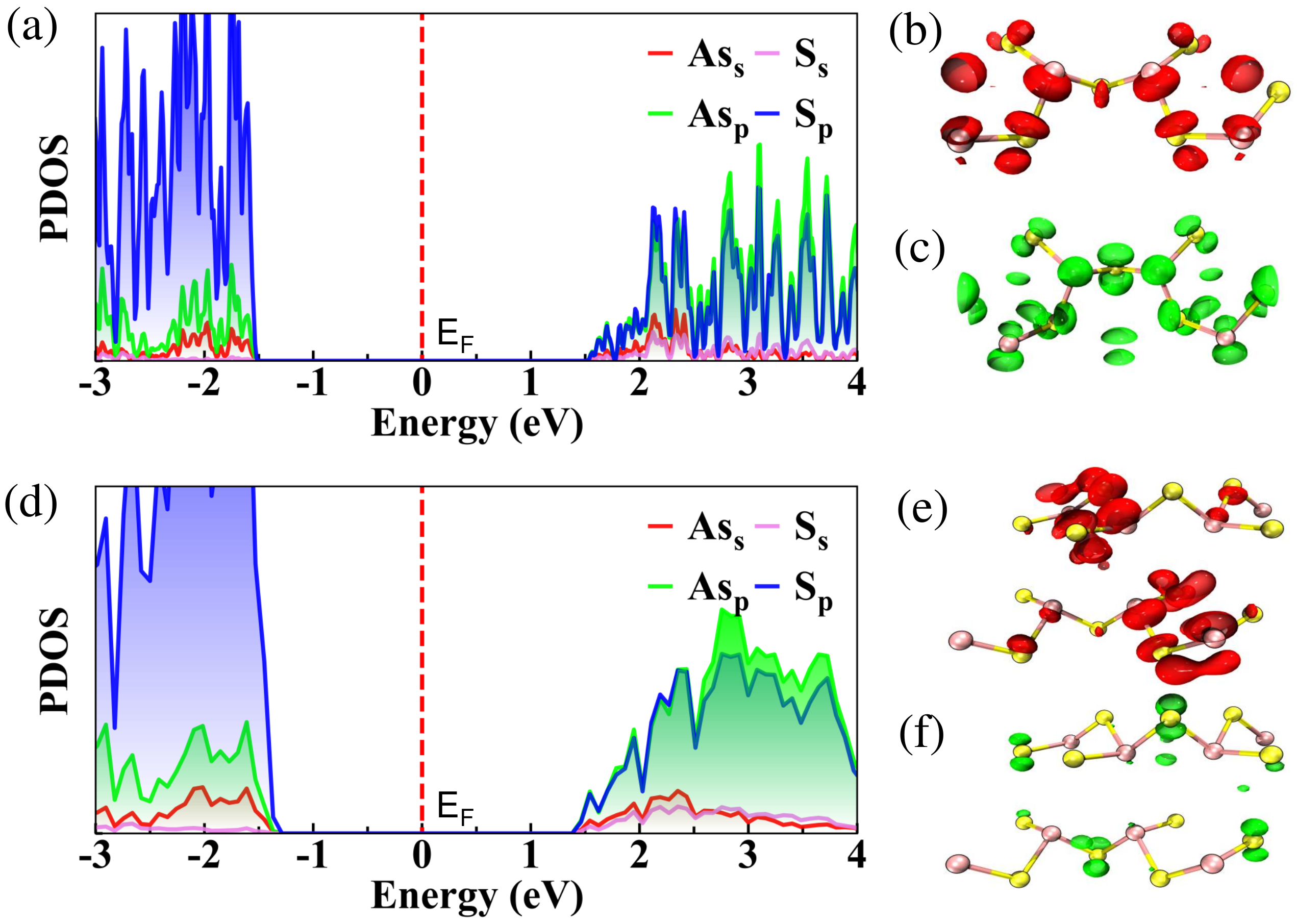}
\caption{(a) The atomic projected density of states (PDOS), %
charge density distributions of (b) CBM and (c) CBM for monolayer
As$_2$S$_3$. (d)--(f) are the corresponding pictures for bilayer
As$_2$S$_3$. The Fermi levels are indicated by red dashed lines. %
}\label{fig3}
\end{figure*}

Besides, we also show the atomic projected density of states (PDOS) %
as well as the charge density distributions of VBM and CBM based on
the HSE06 level for monolayer in Figs.~\ref{fig3}(a) and
\ref{fig3}(c) and bilayer in Figs.~\ref{fig3}(d) and \ref{fig3}(f). %
Evidently, the PDOS and charge density of band edges for monolayer
and bilayer As$_2$S$_3$ show that the VBM is mainly dominated by the
\textit{p} orbital of sulfur atoms with a minor contribution of
antibonding $(As-S)_{in-plane}$ $\sigma^*$ states. On the contrary, %
the CBM is equally contributed by the \textit{p} orbital of sulfur
and arsenic atoms, suggesting a comparable bonding $As-S$ $\pi$ and
anti--bonding $(As-S)_{vertical}$ $\sigma^*$ states. Therefore, we
expect our results for effective masses of electrons and holes will
further unveil the anisotropic electronic transport in As$_2$S$_3$
systems. %

\begin{figure*}[t!]
\includegraphics[width=2.0\columnwidth]{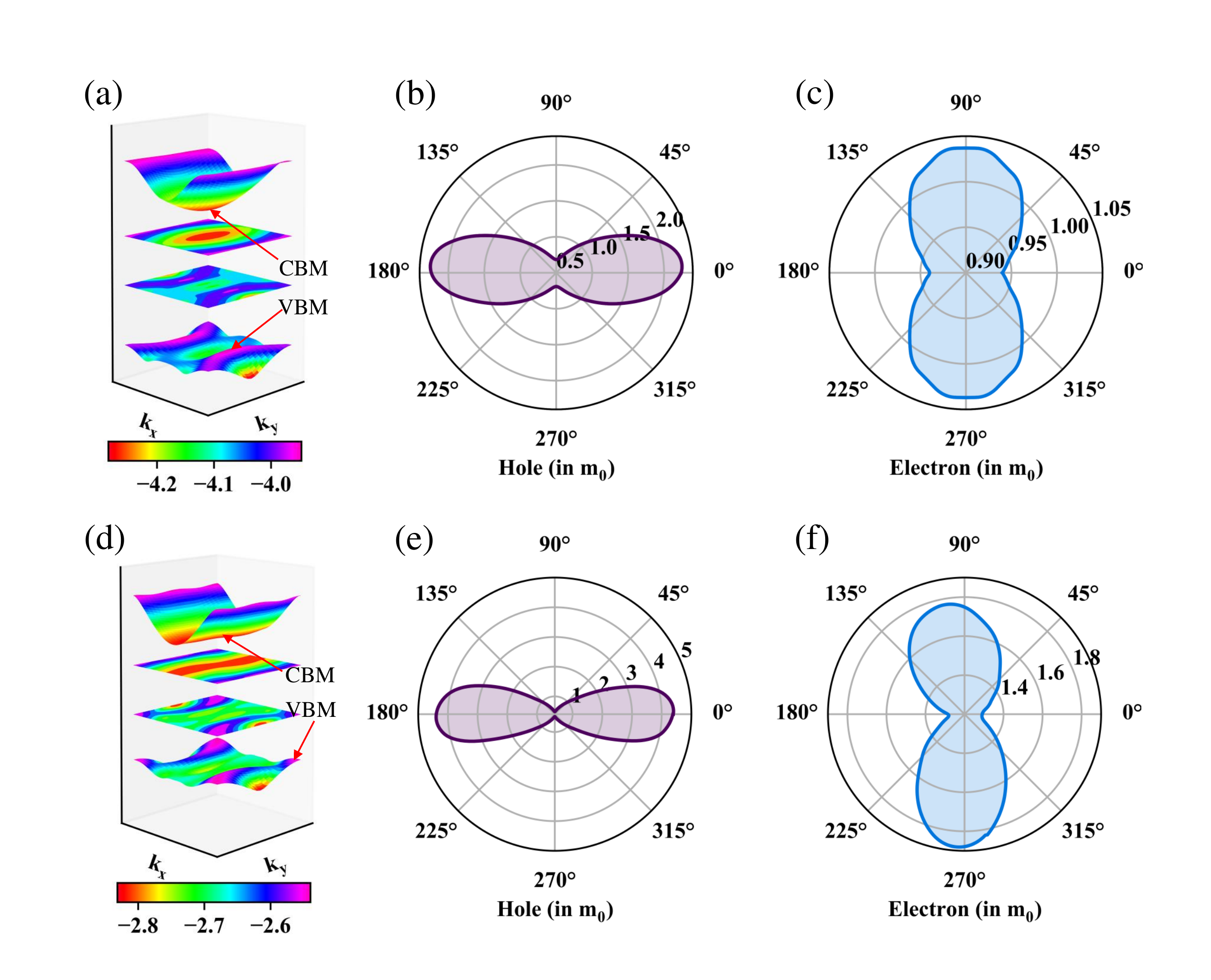}
\caption{(a) Three-dimensional (3D) electronic band structure and
angle-resolved effective masses of (b) holes and (c) electrons
located at VBM and CBM band edges for monolayer As$_2$S$_3$. %
(d)--(f) are the corresponding quantities for bilayer
As$_2$S$_3$. The CBM and VBM are indicated in (a) and (d). %
}\label{fig4}
\end{figure*}

To further confirm the anisotropic electronic transport properties, %
we plot the three-dimensional band structures of monolayer and
bilayer As$_2$S$_3$, shown in Figs.~\ref{fig4}(a) and \ref{fig4}(d), %
respectively. They demonstrate obvious anisotropic nature at the CBM
and VBM. CBM band edges are much more dispersive than that of the
VBM band edges, again indicating the strong anisotropic electronic
transport properties. The effective mass (in units of electron mass
m$_0$) of carriers at band edges can be calculated by
\begin{equation} %
\begin{aligned}
\label{mass}
m^*_i = \hbar^2 [ \frac{\partial^2 E(\textbf{k$_i$}) }{\partial^2 \textbf{k$_i$}} ]^{-1},
\end{aligned}
\end{equation}
where \textit{E}(\textit{\textbf{k$_i$}}) is the electronic energy
dispersion with respect to the electron momentum along \textit{i} %
(a, b, c) direction. During the process of orientation-dependent
effective mass, a uniform \textit{k}--points were sampled along
the radial direction at intervals of 10~degrees to calculate
the band dispersion relation. Then we fitted the dispersion
in a given direction to obtain the angle-dependent effective
masses of holes and electrons~\cite{wang2019vaspkit}. %

Different electronic bands according to the Eq.~(\ref{mass}) will
result in quantitative effective mass of carriers in different
directions. The orientation-dependent effective masses of holes
and electrons are depicted in Figs.~\ref{fig4}(b) and \ref{fig4}(c)
for monolayer and Figs.~\ref{fig4}(e) and \ref{fig4}(f) for bilayer
As$_2$S$_3$ respectively. Obviously, the effective mass along $\vec{a}$
direction is much smaller than that of in the $\vec{b}$ direction. %
As for the holes, an opposite trend is found and the anisotropic
nature of holes is much stronger than the electrons. These
significantly anisotropic transport properties could be beneficial
to the separation of electrons and holes which is highly desired
in the photovoltaic field~\cite{wang20182d}. Furthermore, the
bilayer As$_2$S$_3$ has a larger anisotropic effective masses of
holes and electrons compared with the monolayer situation, which
will lead to asymmetric transport properties, such as the Seebeck
coefficient and electric conductivity and finally potentially
enhance the thermoelectric performance~\cite{dresselhaus2007new,%
heremans2008enhancement}. %

\begin{figure*}[t!]
\includegraphics[width=2.0\columnwidth]{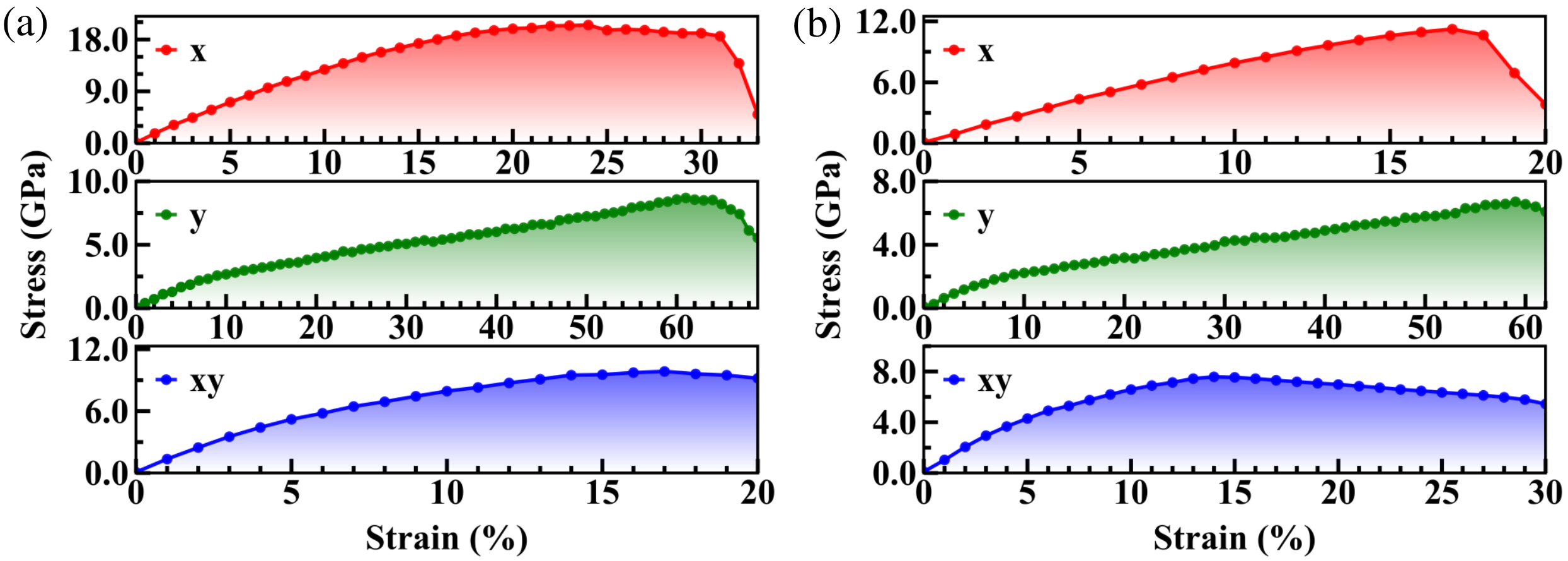}
\caption{The relationship between stress and uniaxial/biaxial
strain for (a) monolayer and (b) bilayer As$_2$S$_3$. In the
calculation of uniaxial strain (for example, in the x
direction), the lattice constant along x-direction is fixed
to a value of \textit{L}$_x$, then the lattice constant
\textit{L}$_y$ along the y-axis and the atom positions are
optimized until reaching the lowest total energy of the
whole system. %
}\label{fig5}
\end{figure*}
%
\subsection{Anisotropic mechanical properties}
The mechanical properties of a material are those properties that
involve a response to an applied strain, which has wide
applications~\cite{greaves2011poisson,gao2017novel,gao2018two}. %
The calculated stress-strain curves of monolayer and bilayer
As$_2$S$_3$ are shown in Figure~\ref{fig5}, which starts a linear
function before the loaded strain is lower than 5\%. Above 5\%, %
monolayer and bilayer As$_2$S$_3$ enter nonlinear (anharmonic)
regions, which are consistent with the previous
work~\cite{mortazavi20202}. Young's modulus \textit{E} is the
slope in the stress-strain curve locating in the linear
region~\cite{gao2017novel}. By fitting calculation, we obtained
the elastic modulus for monolayer (bilayer) are 45.4 (85.1) GPa, %
and 11.3 (27.4) GPa along with the $x$($\vec{a}$) and $y$($\vec{b}$)
directions, respectively. Due to the rectangle crystals of
monolayer and bilayer As$_2$S$_3$, the mechanical properties along
with x-direction can decouple with the y-direction, which further
increases the anisotropic mechanical properties of As$_2$S$_3$
systems. This is completely verified by the Figs.~\ref{fig5}(a)
for monolayer and \ref{fig5}(b) for bilayer As$_2$S$_3$.

For a 2D material, the relationship between the stress $\sigma$, the
in-plane elastic constants tensor $C_{ij}$ ($i,j$=1,2,6) and strain
$\varepsilon$ can be correlated based on the Hooke's law under the
in-plane stress condition~\cite{wang2017lattice,gao2017novel}
\begin{equation}
\label{Eq1}%
\left[
\begin{array}{c}
\sigma_{xx}\\
\sigma_{yy}\\
\sigma_{xy}
\end{array}
\right]=
\left[
\begin{array}{ccc}

C_{11} & C_{12} & 0\\
C_{12} & C_{22} & 0\\
0 & 0 & C_{66}
\end{array}
\right]
\left[
\begin{array}{c}

\varepsilon_{xx}\\
\varepsilon_{yy}\\
2\varepsilon_{xy}
\end{array}
\right]
\end{equation}
here we use the standard Voigt notation which simplifies the tensor
notation into the matrix notation, such as 1--$xx$, 2--$yy$, and
6--$xy$~\cite{andrew2012mechanical}. Besides, since the rectangle
crystals of monolayer and bilayer As$_2$S$_3$, the elastic constants
can be calculated as %
\begin{equation}
  E_S=\frac{1}{2}C_{11}\varepsilon_{1}^2+\frac{1}{2}C_{22}\varepsilon_{2}^2+C_{12}\varepsilon
  _{1}\varepsilon_{2}+2C_{66}\varepsilon_{6}^2
\label{Eq2}%
\end{equation}
where $E_S$ is the strain energy and the tensile strain is defined as
$\varepsilon=\frac{L_i-L_{i0}}{L}$ ($i=x,y$). $L_i$ and $L_{i0}$ are
the strained and unstrained lattice constants along with x- or
y-directions, respectively. To capture the physics, we select the
$\varepsilon_i$ ($i$=1, 2, 6) ranging from the $-2\%$ to $2\%$ with
an increment of 0.5\% to calculate the strain energies under different
strains for all strained structures, including monolayer and bilayer
As$_2$S$_3$. Thus the elastic constants can be obtained by fitting the
Eq.~(\ref{Eq2}).

\begin{figure*}[t!]
\includegraphics[width=2.0\columnwidth]{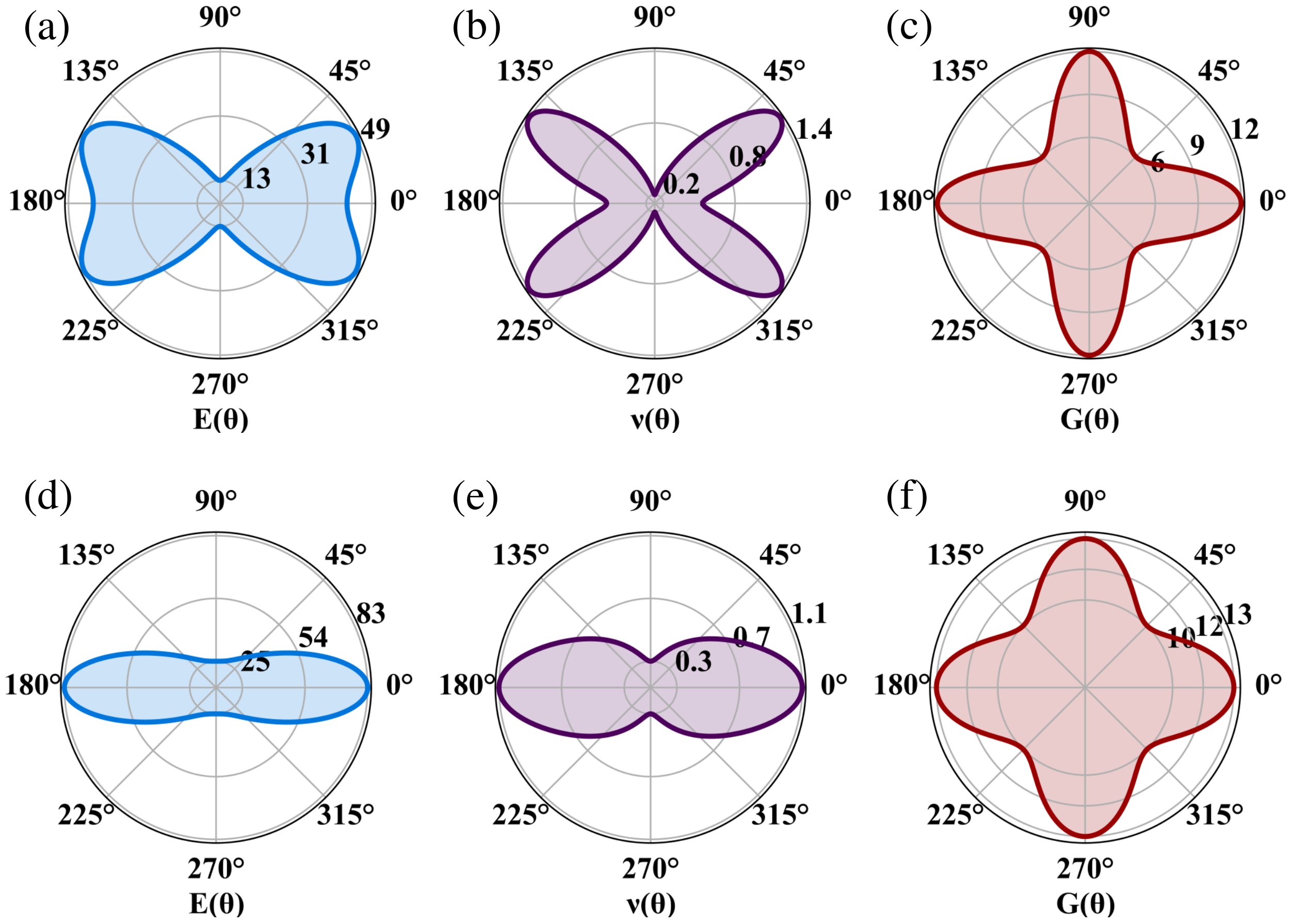}
\caption{The calculated angle-resolved Young's modulus
\textit{E}($\theta$), Poisson's ratio $\nu$($\theta$), and
Shear modulus \textit{G}($\theta$), respectively for (a)--(c)
monolayer and (d)--(f) bilayer As$_2$S$_3$ according to
Eq.~(\ref{Eq3}), (\ref{Eq4}), and (\ref{Eq8}). %
}\label{fig6}
\end{figure*}

To intuitively investigate the mechanical anisotropy of As$_2$S$_3$
systems, we calculate the orientation-dependent Young's modulus
\textit{E}($\theta$), Poisson's ratio $\nu$($\theta$), and Shear
modulus \textit{G}($\theta$) based on%
~\cite{jasiukiewicz2008auxetic,jasiukiewicz2010auxetic}
\begin{equation}
\begin{aligned}
\label{Eq3}
\textit{E}^{-1} = & S_{11} \cos^4 \theta + S_{22} \sin^4 \theta + 2S_{16} \cos^3 \theta \sin\theta  \\
&+2S_{26} \cos \theta \sin^3 \theta  \\
&+( 2S_{12} + S_{66} ) \cos^2 \theta \sin^2 \theta, \\
\end{aligned}
\end{equation}
and
\begin{equation}
\label{Eq4}
\begin{aligned}
-\nu(\theta)/ \textit{E}(\theta) = A + B \cos ( 4 \theta + \psi_1 ), \\
\end{aligned}
\end{equation}
where
\begin{equation}
\label{Eq5}
\begin{aligned}
A = [ ( S_{11} + S_{22} - S_{66} )/2 + 3S_{12} ]/4, \\
\end{aligned}
\end{equation}
\begin{equation}
\label{Eq6}
\begin{aligned}
B = \frac{\sqrt{ (S_{26} - S_{16} )^2 + [S_{12} - (S_{11} + S_{22} - S_{66} )/2]^2 }}{4} \\
\end{aligned}
\end{equation}
\begin{equation}
\label{Eq7}
\begin{aligned}
\tan \psi_1 = \frac{ S_{26}-S_{16} }{S_{12} - ( S_{11} + S_{22} - S_{66} )/2 }  , \\
\end{aligned}
\end{equation}
\begin{equation}
\label{Eq8}
\begin{aligned}
1/ 4 \textit{G} (\theta) = C + D \cos ( 4 \theta + \psi_2 ) , \\
\end{aligned}
\end{equation}
where
\begin{equation}
\label{Eq9}
\begin{aligned}
C = ( S_{11}+S_{22}-2S_{12}+S_{66} )/8 , \\
\end{aligned}
\end{equation}
\begin{equation}
\label{Eq10}
\begin{aligned}
D = \frac{\sqrt{ (S_{66} + 2S_{12} - S_{11} - S_{22} )^2/4 + ( S_{26} - S_{16} )^2 }}{4} \\
\end{aligned}
\end{equation}
\begin{equation}
\label{Eq11}
\begin{aligned}
\tan \psi_2 = \frac{ 2(S_{16} - S_{26}) }{ (S_{66} + 2S_{12} - S_{11} - S_{22}) }. \\
\end{aligned}
\end{equation}
in which $\theta$ $\in$ [0, 2$\pi$] is the conventional angle that
starts from the +$x$ axis corresponding to the $\theta = 0$. In the
experiment, the data of mechanical property sometimes are compliance
constants that have a straightforward relation with elastic tensors: %
$S_{ij}$=$C_{ij}^{-1}$. Our calculated results are presented in
Figs.~\ref{fig6}(a) and \ref{fig6}(c) for monolayer and
Figs.~\ref{fig6}(d) and \ref{fig6}(f) for bilayer As$_2$S$_3$. %

At first glance, for bilayer As$_2$S$_3$, both Young's modulus
\textit{E} and Poisson's ratio $\nu$ decrease to a minimum value
then increase as a function of orientation. The maximum and
minimum values of \textit{E} for bilayer are 83 GPa at $0^{\circ}$ %
($\vec{a}$ axis) and 25 GPa at $90^{\circ}$ ($\vec{b}$ axis). The
corresponding results of $\nu$ are 1.1 and 0.3, respectively. %
Unfortunately, the situation of monolayer As$_2$S$_3$ is
complicated where there exist two maximums for \textit{E} and
$\nu$, separately. As for the Shear modulus \textit{G}, Similar
trends can be found for both monolayer and bilayer As$_2$S$_3$ shown
in Figs.~\ref{fig6}(c) and \ref{fig6}(f). The maximum \textit{G}
is around 13 GPa for bilayer, while the minimum \textit{G} is
10.3 GPa and 5.9 GPa for monolayer and bilayer As$_2$S$_3$, %
respectively.

These results suggest that the anisotropic mechanical properties
are obvious both in monolayer and bilayer As$_2$S$_3$. What is more, %
the outcome from the strain--energy method is also verified by
Figure~\ref{fig5}, which confirms the correctness and
consistency of our computational methods. %
%
%
In our present work, the anisotropic factor of monolayer and bilayer
are 3.15 and 3.32, respectively, which is relatively good agreement
with the previous experimental measurement~\cite{kins2019highly}
and is quite larger than the renowned black phosphorous with
with experimentally confirmed and an anisotropic factor
of 2~\cite{tao2015mechanical}. %

Furthermore, the absolute value of Young's modulus increases from
the monolayer in Figure~\ref{fig6}(a) to the bilayer
Figure~\ref{fig6}(d). Our results in $\vec{a}$-direction are
consistent with the experimental values, but a little smaller
than that of the experimental values in $\vec{b}$-direction. This
limited discrepancy is probably derived from the layer-dependent
effect. In the experiment, the investigated samples are more two
layers of As$_2$S$_3$, while our calculation is exactly the
bilayer system~\cite{kins2019highly}. However, both the
experimental data and our calculated results confirm the high
anisotropic mechanical properties of 2D As$_2$S$_3$ material. %

We find that the highly anisotropic mechanical properties in
As$_2$S$_3$ systems can be explained by As-S-As bond angles along
$\vec{a}$-direction. All optimized $As-S$ bond lengths range
from 2.28 to 2.31 {\AA} in both monolayer and bilayer As$_2$S$_3$, %
but they have different bond angles. We find that all the
As-S-As bond angles can be classified into two types, one is the
mainly elongated bonds along with $\vec{a}$-direction, called
A-type shown in Figs.~\ref{fig1}(a) and \ref{fig1}(b), with
A-type bond angles of $89^{\circ}$ and $88^{\circ}$, respectively. %
In contrast, the other As-S-As bond angle is called B-type with
a value of $102^{\circ}$ and $101^{\circ}$ for monolayer and
bilayer As$_2$S$_3$, separately. The smaller As-S-As bond angle
of A-type along $\vec{a}$-direction will enhance the strength
of the As-As bonds ($As_1-As_2$ in Figure~\ref{fig1}(a)) and
S-S bonds ($S_1-S_2$ in Figure~\ref{fig1}(a)). However, the
bond angles of the B-type result in the weakness of As-As
bonds ($As_1-As_3$ in Figure~\ref{fig1}(a)) and S-S bonds %
($S_2-S_3$ in Figure~\ref{fig1}(a)) along $\vec{b}$-direction. %
As a consequence, large anisotropic Young's modulus \textit{E}
and $\nu$ are identified and verified. %

\section*{Conclusions}
In this work, we have systematically explored the charge
distribution, electronic band structures, angle-resolved
effective masses, strain-stress curves, %
orientation-dependent Young's modulus, Poisson's ratio, %
and Shear modulus for monolayer and bilayer As$_2$S$_3$ by
first-principles calculations. The result shows that
monolayer and bilayer As$_2$S$_3$ have significantly large
anisotropy of electronic and mechanical properties. The
electronic anisotropy would make 2D As$_2$S$_3$ a superior
candidate for applications in the photovoltaic field where
the generated holes and electrons need to be separated. %
More interestingly, the calculated anisotropic factor of
monolayer and bilayer As$_2$S$_3$ are 3.15 and 3.32, %
respectively, which are quite larger than the renowned
black phosphorous with experimentally confirmed and an
anisotropic factor of 2. We expect our study will
provide an effective route to flexible
orientation-dependent nanoelectronics, nanomechanics, %
and has implications in promoting related experimental
investigations. %


\quad\\
{\noindent\bf Author Information}\\


{\noindent\bf ORCID}\\
Xuefei Liu: 0000-0003-0154-474X \\
Zhaofu Zhang: 0000-0002-1406-1256 \\
Zhibin Gao: 0000-0002-6843-381X \\

{\noindent\bf Notes}\\
The authors declare no competing financial interest.


\begin{acknowledgement}\\
This work is supported by the National Natural Science Foundation of
China (Grant No. 61564002 and 11664005); the Joint Foundation of
Guizhou Normal University (Grant No. 7341); Scientific and
Technological Cooperation Projects of Guizhou Province, China %
(Grant No. [2013] 7019). Science and technology planning project
of Guizhou province (Grant No. 2017-5736-009). J.-S. Wang and
Z. Gao acknowledge the financial support from MOE tier 1 funding
of Singapore (grant no. R-144-000-402-114). %
\end{acknowledgement}


\providecommand{\latin}[1]{#1}
\makeatletter
\providecommand{\doi}
  {\begingroup\let\do\@makeother\dospecials
  \catcode`\{=1 \catcode`\}=2 \doi@aux}
\providecommand{\doi@aux}[1]{\endgroup\texttt{#1}}
\makeatother
\providecommand*\mcitethebibliography{\thebibliography}
\csname @ifundefined\endcsname{endmcitethebibliography}
  {\let\endmcitethebibliography\endthebibliography}{}

\end{document}